\documentclass{svjour2}                     
\smartqed  
\usepackage{graphicx}
\usepackage{bm}
\usepackage{color}
\usepackage{amsmath}

\journalname{J. Stat. Phys.}

\newcommand{\bea}{\begin{eqnarray}}
\newcommand{\eea}{\end{eqnarray}}
\newcommand{\nn}{\nonumber}

\newcommand{\la}{\langle}
\newcommand{\ra}{\rangle}

\newcommand{\f}{\frac}

\newcommand{\agedistr}{\phi(t)}
\newcommand{\tx}{t_x}
\newcommand{\tmax}{t_{\mbox{\tiny{max}}}}
\newcommand{\concx}{{{p}}}
\newcommand{\concp}{{{P}}}

\newcommand{\vexp}{V_{\mbox{\tiny{exp}}}(t)}
\newcommand{\vlin}{V_{\mbox{\tiny{lin}}}(t)}
\newcommand{\xexp}{\concx_{\mbox{\tiny{exp}}}(t)} 
\newcommand{\xlin}{\concx_{\mbox{\tiny{lin}}}(t)} 
\newcommand{\xxexp}{\concx_{\mbox{\tiny{exp}}}}
\newcommand{\xxlin}{\concx_{\mbox{\tiny{lin}}}}

\newcommand{\etaxexp}{\eta^2_{\concx_{\mbox{\tiny{exp}}}}}
\newcommand{\etaxlin}{\eta^2_{\concx_{\mbox{\tiny{lin}}}}}

\begin{document}

\title{Deterministic and stochastic descriptions of gene expression dynamics}


\titlerunning{Descriptions of gene expression dynamics}        

\author{Rahul Marathe*, Veronika Bierbaum*, David Gomez, and Stefan Klumpp}

\authorrunning{R. Marathe, V. Bierbaum, D. Gomez, and S. Klumpp} 

\institute{R. Marathe, V. Bierbaum, D. Gomez, S. Klumpp \at
               Max Planck Institute of Colloids and Interfaces, Science Park Golm, 
14424 Potsdam, Germany\\
\email{rahul.marathe@mpikg.mpg.de}     \and
D. Gomez \at 
Department of Physics, Freie Universit{\"a}t Berlin, Arnimallee 14,  14195 Berlin, Germany      \and
* R. M. and V. B. contributed equally.
}

\date{Received: date / Accepted: date    }

\maketitle

\begin{abstract}
A key goal of systems biology is the predictive mathematical description of gene regulatory circuits. Different approaches 
are used such as deterministic and stochastic models, models that describe cell growth and division explicitly or implicitly etc. Here we consider 
simple systems of unregulated (constitutive) gene expression and compare different mathematical  descriptions systematically to obtain insight into the errors that are introduced by 
various common approximations such as describing cell growth and division by an effective protein degradation term. In particular, we show that the 
population average of protein content of a cell exhibits a subtle dependence on the dynamics of growth and division, the specific model for volume 
growth and the age structure of the population. Nevertheless, the error made by models with implicit cell growth and division is quite small. 
Furthermore, we compare various models that are partially stochastic to investigate the impact of different sources of (intrinsic) noise. This 
comparison indicates that different sources of noise (protein synthesis, partitioning in cell division) contribute comparable amounts of noise if 
protein synthesis is not or only weakly bursty. If protein synthesis is very bursty, the burstiness is the dominant noise source, independent of 
other details of the model. Finally, we discuss two sources of extrinsic noise: cell-to-cell variations in protein content due to cells being 
at different stages in the division cycles, which we show to be small (for the protein concentration and, surprisingly, also for the protein copy 
number per cell) and fluctuations in the growth rate, which can have a significant impact.
\end{abstract}


\section{Introduction}
\label{intro}

With the emergence of systems biology and synthetic biology, concepts and methods from mathematics, physics and engineering are increasingly used 
in the life sciences \cite{Alon03,Csete02,Andrianantoandro06}. In particular, two central goals of this field are to predict the dynamics of gene 
expression based on mathematical descriptions of the genetic networks of a cell and to design genetic circuitry based on well-characterized regulatory 
elements \cite{Hasty,Bintu2005,Guido06,Wall}. The progress of this research program has however also highlighted a number of generic complications that arise from 
the fact that all genetic circuits function in  a cellular chassis that itself is dynamic and adapts to external conditions, which can have unexpected 
effects on circuit function \cite{Klumpp09,TanYou,Scott,KlumppPlasmid,Kwok}. This observation raises the question what mathematical description is appropriate 
for the description of genetic circuitry in a dynamic cell. In particular, even if the external conditions are constant and the cells exhibit 
'balanced growth' {\color{black}(a  steady state of all cellular parameters except for the overall  exponential growth of the culture)}, each individual cell grows and 
divides and, while doing so, doubles its content of all cellular components. Some of the components will clearly affect the function of any gene circuit,  
the most important example being the duplication of the circuit genes themselves. In mathematical models of genetic circuits, these effects are often ignored 
and described by an average gene copy number and an effective degradation of the protein that mimics the dilution of a protein concentration due to cell growth 
in the absence of its synthesis. In this article, we therefore ask how strongly cell growth within the division cycle affects gene expression and whether models 
that do not describe growth and division explicitly introduce big errors through that approximation. 

Another facet of the question which mathematical description to use is the question whether such a description should be deterministic or stochastic. 
It has been realized in recent years that often the relevant molecules are present in the cell in low copy numbers, giving rise to large fluctuations 
and thus requiring a stochastic description of gene expression  \cite{AdamsPNAS97,ElowitzSci02,GuptaBio95,Rao02,Kaern05}. 
The foundations for this view have been laid long ago \cite{SpudichNat76,BergJTB78}, but the progress in single-molecule and single-cell technology 
{\color{black} now allows the direct observation of these effects and the quantification of fluctuations from time series or from cell-to-cell variability} \cite{YuSci06,xie,Golding,ElowitzSci02}. Stochasticity in gene expression has been studied extensively 
from a theoretical point of view, see e.g.\ \cite{BergJTB78,KoJTB91,YcartTPB95,CookPNAS95,BergBiopJ2000,HastyPNAS2000,OudenaardenPNAS01,SwainPNAS2002,PaulssonReview,Scott07}. 
Here we ask about the sources of stochasticity, as noise can be generated at many points in the process of protein synthesis and by the partitioning during 
cell division. Many of the noise sources have been studied before, but we are interested in a systematic comparison of their impact. Specifically, we ask 
whether there is a dominant source of noise, and whether the noise predicted from models with explicit cell growth and division differs from what is obtained 
from implicit cell division models. 

It turns out that the question of stochasticity and the dependence of gene expression on the growth and division cycle are closely related:  
{\color{black} The variation of  a protein concentration during the division cycle is observed as a cell-to-cell variation in that concentration in snapshots of cell cultures (where the division cycles of different cells are typically not synchronized, i.e.\ different cells divide at different times).} 
We therefore 
also determine the effective 'noise' that arises from the dependence on the division cycle (which in fact is a deterministic component of the observed 'noise' and 
is seen as part of the so-called 'extrinsic noise' that is common to different genes \cite{ElowitzSci02,SwainPNAS2002}). 

The paper is organized as follows: We start with deterministic descriptions of gene expression in section \ref{sec:1}, where we discuss the effects of the division 
cycle and approximations that 'average out' the division cycle. In sections \ref{sec:noiseI} and \ref{sec:bursts} we discuss several simple models that describe various 
processes of gene expression stochastically to address the question of the relative importance of various sources of stochasticity. We derive analytical results for some key characteristics of the noise.  Here we focus on intrinsic noise, i.e. 
noise inherent in the synthesis and division process and specific to one gene. Extrinsic noise is discussed in section \ref{sec:extrinsic}, where we come back to the 
dependence of protein concentrations on the division cycle and show that the effective 'noise' resulting from this dependence is small (section \ref{sec:6}). In addition, 
we also include a discussion of fluctuations of the growth rate (section \ref{sec:growth}).  We end with some general conclusions in section \ref{sec:concl}, {\color{black} where we summarize the relative importance of various sources of noise and cell-to-cell variations and discuss the minimal ingredients {\color{black}to arrive at} realistic descriptions of gene expression.}

\section{Deterministic descriptions of gene expression}
\label{sec:1}

\subsection{Basic model}
\label{sec:modelBasics}

We will start by discussing a simple deterministic model of protein synthesis that accounts for the effects of the cell division cycle, 
specifically cell division itself and gene duplication, onto protein synthesis. Living cells grow and divide, while in the meantime, 
proteins are continuously synthesized inside the cell.  We determine the amount of protein synthesized within a cell cycle and the 
corresponding concentrations for both exponential and linear cell growth. 

The number of copies of a specific protein in a cell, $\concp(t)$, is described by the following dynamics:
\begin{eqnarray}\label{synthesis}
\dot{\concp}=\alpha g -\beta \concp,
\end{eqnarray}
where $\alpha$ is the protein synthesis rate, $g$ is the gene copy number and $\beta$ is the protein degradation rate {\color{black}(typical parameter values are summarized in appendix \ref{app:param})}. Throughout this work, we will assume that the proteins are stable ($\beta\approx 0$), as it is typically the case for bacterial proteins \cite{NathJBC70}.

While the proteins are synthesized, the cell also grows and divides. Divisions take place at integer multiples of the doubling time  $T$. 
Here we treat cell division as a deterministic process that occurs instantaneously. At the time of division, the amount of our protein of interest 
is divided equally among two daughter cells, so that its amount per cell is simply divided by 2. The same partitioning applies to all other contents 
of the cell, and therefore, in a steady state of growth,  all content of the cell has to be doubled between divisions. Specifically, we are interested in 
the doubling of the gene that encodes our protein of interest. This gene, which we assume to be present as a single copy in the genome of the cell, 
is doubled at a time $t_x$ after the last division (and, of course, divided by 2 at the time of division). Therefore, the gene copy number $g$ that enters 
Eq. (\ref{synthesis}) is given by $g=1$ for times $0\leq t<t_x$ after division and by $g=2$ for times $t_x\leq t<T$. Another important characteristic of the 
cells that has to double over the doubling time is the total cell mass or the cell volume. We will come back to this point below, when we discuss the 
concentration of the protein.

\begin{figure}[tb]
\begin{centering}
\includegraphics[width=.7\textwidth]{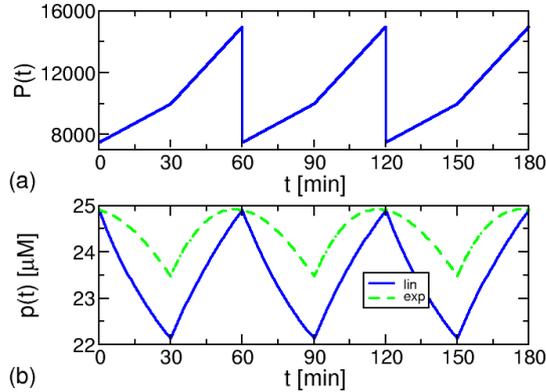}
\caption{Variation of the protein number $\concp(t)$ (a) and concentrations $\xlin$ and $\xexp$ (b) over the cell division cycle. (a) The protein copy number increases from $\concp_0=7500$ to $2\concp_0=15000$ during a cell division cycle. Note that the protein synthesis rate doubles at time $\tx$ after each cell division, where the gene is replicated. 
(b) The corresponding protein concentration decreases transiently during the division cycle. This effects is more pronounced for linear volume growth (solid blue line) than for exponential volume growth (green dashed line) within the division cycle.
The parameters are $\alpha=5000/T$, $T=60$ min, $\tx=30$ min, $V_0=0.5 \mu$m$^3$. \label{fig1}}
\end{centering}
\end{figure}

Now we consider our gene to be in a 'steady state' of protein synthesis, in the sense that the protein level only depends on the time in the 
division cycle, but is the same if corresponding time points in different cycles are compared. In that case, the protein copy number at the end 
of the cycle is exactly twice that at its beginning, i.e. $P(t=T)=2P_0=2P(t=0)$ (here and in the following, we measure time with respect to the time 
of division, i.e. assume that divisions take place at integer multiples of $T$). This condition, which can be considered as a singular boundary condition 
for Eq. \ref{synthesis} with times restricted to the interval $[0, T]$, determines the time course of the copy number of our protein of interest per cell, 
%
%
\begin{equation}
 \concp(t) = \begin{cases} 
      \alpha (t + 2T - t_x) \qquad  {\rm for}\quad 0\le t \le t_x \\ 
    2 \alpha (t + T - t_x)  \qquad {\rm for}\quad t_x < t \le T. 
  \end{cases}
\end{equation}
Immediately after division, there are $P_0=\alpha(2T-t_x)$ copies of the protein in the cell, and the same number is synthesized over the doubling time $T$ (Figure \ref{fig1}). This synthesis occurs in two phases, from one or two copies of the gene, respectively. One can define an effective synthesis rate $\alpha_{\rm eff}=\alpha(2-t_x/T)$, then the number of proteins synthesized over the division cycle has the intuitive form $\alpha_{\rm eff}T$. 

We now turn to the corresponding concentration of the protein. This will be denoted by $p$ and is given by $p=P/V$, the number of protein molecules per cell divided by the cell volume $V$. It therefore also depends on the time course of the cell volume over the division cycle. The functional form of that time dependence has been debated for a long time, see for example a recent discussion in ref. \cite{Cooper}.  Here we use two models that have been proposed, namely linear and exponential growth of the cell volume,  which we indicate by the subscripts 'lin' and 'exp', respectively. 

We denote the  cell volume at the beginning of a cycle by $V(t=0)\equiv V_0$. In a steady state of growth, this volume must have doubled at the end of the cell cycle, such that $V(t=T)=2V_0$. Using this constraint, the cell volume $V(t)$ is given by
\begin{eqnarray}
\vlin&=&V_0(1+t/T),
\end{eqnarray}
for linear and by
\begin{eqnarray}
 \vexp&=&V_0\exp{\left(\frac{\ln 2~t}{T}\right)}
\end{eqnarray}
for exponential growth. As a consequence, the concentration of our protein at the beginning and at the end of a division cycle is equal, $p(t=0)=p(t=T)=P_0/V_0$. However, it decreases between divisions as the protein copy number initially grows more slowly than the volume. When the gene is duplicated, the protein copy number growth speeds up and becomes faster than volume growth and the concentration increases for times $t_x<t<T$ such that the concentration returns to its initial value. This temporary decrease of the protein concentration is more pronounced for linear  than for exponential volume growth, as can be seen in Fig.  \ref{fig1}(b).  The extent of this decrease depends on the timing of gene duplication (which is dependent on the position of the gene with respect to the origin of DNA replication \cite{CHelmstetter,Bremer1996}). For example, in the extreme case, where the gene is duplicated immediately after or before cell division, the protein content increases approximately linearly, and thus, for linear volume growth, the concentration is almost constant over the division cycle. We will come back to this point in section \ref{sec:6}, when we discuss the contribution of  division cycle effects to the observed 'noise' in the protein content.

%
%

\subsection{Population averages}

The dynamics described so far is observable in experiments that track the content of specific proteins in single cells. Such experiments have been done (e.g.,\  \cite{YuSci06,xie,cookson}), although most of these studies were more focused on stochastic effects.  In many experiments, however, what is observed is the population average of the protein content per cell. Unless the cell culture is specifically  prepared  to synchronize the division cycles of these cells, the population will consist of many cells ($\sim 10^9$ in a typical bacterial culture) that divide in an asynchronous fashion. Averages of cellular properties over such populations will in general not only depend on the dynamics of the observable over the division cycle, but also on the age distribution in the population, i.e. the distribution of the time points in the division cycle at which these cells are. The latter depends on the experimental setup. We consider two cases, an exponential and a constant age distribution. The exponential age distribution,  %
\begin{eqnarray}\label{age_distr}
\agedistr=\frac{2\ln 2}{T}\exp{\left(-\frac{\ln 2~t}{T}\right)},
\end{eqnarray}
applies to asynchronous cultures with an exponentially growing population size, where there are more young cells than old cells. The average age of a cell in such culture is $\langle{t}\rangle=\int^T_0 t~\agedistr~dt=T (1/\ln 2-1)\approx 0.44 T$.

In addition we consider a constant age distribution, $\phi(t)=1/T$, which is obtained if, for example, after each cell division only one of the daughter cells is kept and analyzed. An example of such an experimental setup is the 'mother machine' that was described recently \cite{Wang__Jun2010}.



The protein copy number per cell, averaged over such an exponentially growing population, is given by
\begin{equation}\label{avNo_expAge}
\langle \concp\rangle=\int_0^T \concp(t)\phi(t)dt= \frac{\alpha  T  2^{1-\tx/T} }{\ln 2}.  
\end{equation}
This result can be rewritten as  $\langle \concp \rangle= \alpha \langle g \rangle/ \beta_{\rm eff}$ with an effective degradation rate of the protein $\beta_{\rm eff}=(\ln 2)/T$ that describes the loss of proteins due to cell division (with half-life equal to the doubling time of the cells),  and the average copy number of the gene $\langle g \rangle= 2^{1-\tx/T})$.  
For comparison, the average protein copy number per cell  in a population with constant age distribution is $\langle \concp \rangle=\alpha T [3- 2\tx/T + \frac{1}{2}(\tx/T)^2]$. Notice that this is in general not equal to $3P_0/2$. The numerical comparison with Eq.\ (\ref{avNo_expAge}) shows that the average protein number is approximately 4 \% larger with the constant age distribution than with the exponential age distribution. 

The average concentration can be calculated in the same way, but is more involved due to the age-dependence of the volume. We give only the result for exponential volume growth and an exponential age distribution. In this case, we obtain
\begin{eqnarray}
\langle \concx \rangle&=& \int_0^T \concx(t)\phi(t)dt=\frac{\alpha T}{V_0} \times \frac{1/2+2^{-2\tx/T} +2 \ln 2-  \ln 2\ \tx/T}{2 \ln 2}. 
\end{eqnarray}
This can be compared to the  'mean field'  result $\langle \concx(t) \rangle\simeq\langle P \rangle / \langle V \rangle$ that is obtained from the average protein number and the average volume. Using $\la V\ra=(2 \ln 2)V_0$, that approximation leads to
\begin{equation}\label{MF}
\la p\ra \simeq \frac{\alpha  T\,  2^{1-\tx/T} }{2 (ln 2)^2 V_0}\simeq 1.04 \frac{\alpha T \la g \ra}{V_0}.
\end{equation}
A numerical comparison with the exact result shows that they differ by less than 0.3 \% for all values of the replication time $\tx$. Likewise, we find that the average concentrations for linear and exponential volume growth also differ only by a few percent.

\subsection{Averaging out the cell division cycle}

The observation that the 'mean field' approximation for the protein concentration given in Eq.\ \ref{MF} is rather accurate suggests that the dynamics on time scales that are longer than the generation time can actually be described by the following equation
\begin{equation}\label{eq:LT}
\dot{\concx}=\frac{\alpha \la g\ra}{\la V \ra} -\beta_{\rm eff} \concx,
\end{equation}
with $\beta_{\rm eff}=\ln 2/T$ as before (or $\beta_{\rm eff}=\beta+\ln 2/T$, if the protein is unstable). The equation can also be interpreted as describing the dynamics of the average concentration in a population of non-synchronized cells. Through $\beta_{\rm eff}$, the equation describes the loss of protein due to growth and division of the cells as an effective degradation. As protein concentration is actually diluted out by volume growth throughout the division cycle (in contrast to the protein number per cell, which experiences dilution through instantaneous reduction by 50 \% at division), and thus its variations through the cycle are relatively small (Fig. \ref{fig1}b), this approximation can be expected to be quite good. The same approximation can also be used for the average protein copy number per cell, but there one has to keep in mind that variations over the division cycle that are neglected, are stronger, as the protein number $P$ varies 2-fold over the cycle.

\subsection{A remark on messenger RNA}
\label{sec:mRNA}

Protein synthesis is a process that occurs in two steps, transcription and translation. In the first step, the gene sequence is copied into a mRNA, which subsequently serves as a template for protein synthesis. A more complete description of the process thus describes the copy numbers of the protein ($P$) and of the mRNA ($M$),
\begin{eqnarray}
\dot{M} & = & \alpha_m g -\beta_m M \nn \\
\dot{P} & = & \alpha_pM-\beta_p P, \label{eq:P_M}
\end{eqnarray}
{\color{black} with $\alpha_m,~\alpha_p$ and $\beta_m$, $\beta_p$ being the growth and degradation rates of mRNA and protein, respectively.}
In many cases, however, mRNA is rather short-lived and one can approximate the equation for $M$ by its steady state, $M=\alpha_m g/\beta_m$. In that case, we are back to Eq.  \ref{synthesis} with $\alpha=\alpha_p\alpha_m/\beta_m$. 

This approximation is specifically suited for gene expression in bacteria, where typically mRNA lifetimes are of the order of a few minutes \cite{Bernstein,xie}, while proteins, as mentioned above, are mostly stable \cite{NathJBC70,ReehMGG79}. This means that when a gene is turned off and synthesis of the corresponding mRNA and protein is stopped, the mRNA will disappear with a half-life  of a few minutes, while the protein is diluted out through cell growth and division and its half-life is given by the doubling time, which is typically of the order of 1 hour (the range for E. coli is 20 min -- many hours).


\section{Sources of (intrinsic) stochasticity}
\label{sec:noiseI}

As mentioned in the introduction, the copy numbers of some proteins can be small, so that fluctuations play an important role, and stochastic descriptions of the dynamics of gene expression are required. In general, all steps in the synthesis pathway of proteins are stochastic processes. The same is true for the degradation of the protein if that protein is unstable. In addition, the partitioning of the copies of that protein during cell division also adds to the noise. We will now consider these different  sources of noise separately {\color{black}
to characterize the noise arising from different sources  in a systematic way.}\footnote{There are some sources of noise that are specific to particular situations, e.g.\ to highly transcribed genes with dense traffic of RNA polymerases \cite{Klumpp2008b,Klumpp2011}. These will not be considered here.} 
{\color{black}
In these considerations, we aim at understanding the relative importance of  different sources of stochasticity rather than at accurately capturing the complicated processes that govern protein 
production in precise biological detail. Specifically we ask which sources contribute to the noise level observed in the protein number and whether there is a dominant source. In this sense, the most realistic 
model is the one that includes stochastic effects in all processes considered here, but we are interested in whether a reduced model may be sufficient. 

We use a bottom-up approach to study the contributions from cell division, protein synthesis, and finally transcription and translation. We start with a stochastic version of the models described in section \ref{sec:modelBasics}, i.e. with models that treat protein synthesis as a simple one-step process. Effects that are due to the two-step nature of protein synthesis (transcription and translation) will be discussed later in section \ref{sec:bursts}.
The most basic model thus describes  protein synthesis and cell division, and we study three version of this scenario. First, we take the partitioning of proteins into daughter cells upon division to be stochastic (section 3.1), but describe protein 
synthesis deterministically. Second, we treat protein synthesis as a stochastic process but partitioning during cell division as deterministic (section 3.2). Finally, both synthesis and cell 
division are considered as stochastic processes (section 3.3). Our analysis shows that the two noise sources contribute similarly to the overall noise, so none of the noise source is dominant. 

In section \ref{sec:bursts}, we discuss models that explicitly treat protein synthesis as occurring in two steps, transcription and translation. The resulting noise is then characterized in terms of a parameter termed 'burst size', that characterizes the average number of proteins synthesized per mRNA copy. Here, high burstiness leads to a significant increase in the noise with bursty protein synthesis then being the dominant source of stochasticity. Thus, under the conditions of high burstiness, therefore a reduced model that neglects other sources of noise can provide a realistic description of the dynamics.}

All the sources of stochasticity we discuss here produce so-called intrinsic noise \cite{ElowitzSci02}, i.e.\ the fluctuations are specific to the gene/protein under consideration and the fluctuations in the level of two different proteins are uncorrelated. Sources of extrinsic noise, which affects all genes will be discussed in section \ref{sec:extrinsic}. 

\subsection{Stochastic partitioning during cell division}

We first consider the case, where protein synthesis is described by a deterministic process, but where proteins are distributed stochastically into the daughter cells during cell division. Specifically, we consider the case, where each copy of the protein has probability $r=1/2$ to end in each of the two daughter cells. This means that in every generation a constant number $Q=\alpha T$ of proteins is newly synthesized, but the initial copy number of the protein at the beginning of the division cycle fluctuates due to the stochastic partitioning during cell division. Fig. \ref{fig:stoch}(a) shows a time series of such a process as obtained from simulations. 

For this case, a number of characteristics can be obtained analytically {\color{black} using a method described in ref.\ \cite{BrennerPRL07}, which we summarize briefly in appendix \ref{app:method}}. For example, the average copy number of the protein directly after 
cell division is $\la P_0 \ra= Q=\alpha T$ and the variance of that number is $\delta P_0^2=2Q/3$. Two commonly used characteristics of noise are the noise strength $\eta^2$ defined as
\begin{equation}
\label{def_eta}
\eta^2=\frac{\la (P-\la P\ra)^2\ra}{\la P \ra^2}
\end{equation}
and the Fano factor $F=\eta^2 \la P\ra$.  $\eta^2$ typically scales as $\eta^2\sim 1/\la P\ra$, so the latter parameter provides a characterization of the prefactor of that scaling. 
In our specific case, we obtain 
\begin{equation}\label{eta:stochPart}
\eta_0^2=\f{2}{3\la P_0\ra}
\end{equation}
or $F_0=2/3$ (the index '0' in these expressions indicates that we have taken averages over a population of cells immediately after division), plotted in Figure  \ref{fig:stoch}(d).

\subsection{Stochastic protein synthesis}

Next we consider the stochasticity that is inherent in the protein synthesis process itself. To disentangle it from the effects of stochastic partitioning we first describe partitioning deterministically, i.e.\ we consider the case where each daughter cell inherits exactly one half of the protein molecules (Figure \ref{fig:stoch}b). 

We consider again one lineage of cells. Between two cell divisions proteins are synthesized stochastically with rate $\alpha$. At the time of cell division (integer multiples of the doubling time $T$), the protein number is divided by two (if the protein number $P$ is an odd number, we take the number after division to be either $(P+1)/2$ or $(P-1)/2$, each with
probability $1/2$, so strictly speaking, there is a minimal remnant of stochasticity in our deterministic description of division as well). 
To keep the discussion simple, we assume here that the synthesis rate is constant, i.e.\ we neglect the fact that the synthesis rate changes upon duplication of the gene. 
We find 
\begin{equation}
\la P_0\ra=\alpha T,\qquad \delta P_0^2=\f{\alpha T}{3}\qquad{\rm{and}}\qquad \eta_0^2=\f{1}{3\la P_0\ra}.
\end{equation}
The last result implies that the  Fano factor is $F_0=1/3$, which is just half of what we have seen for stochastic partitioning (Eq. \ref{eta:stochPart}).

\subsection{Both sources combined}
\label{sec:combined}

\begin{figure}
\vspace{0.5cm}
\includegraphics[width=\textwidth]{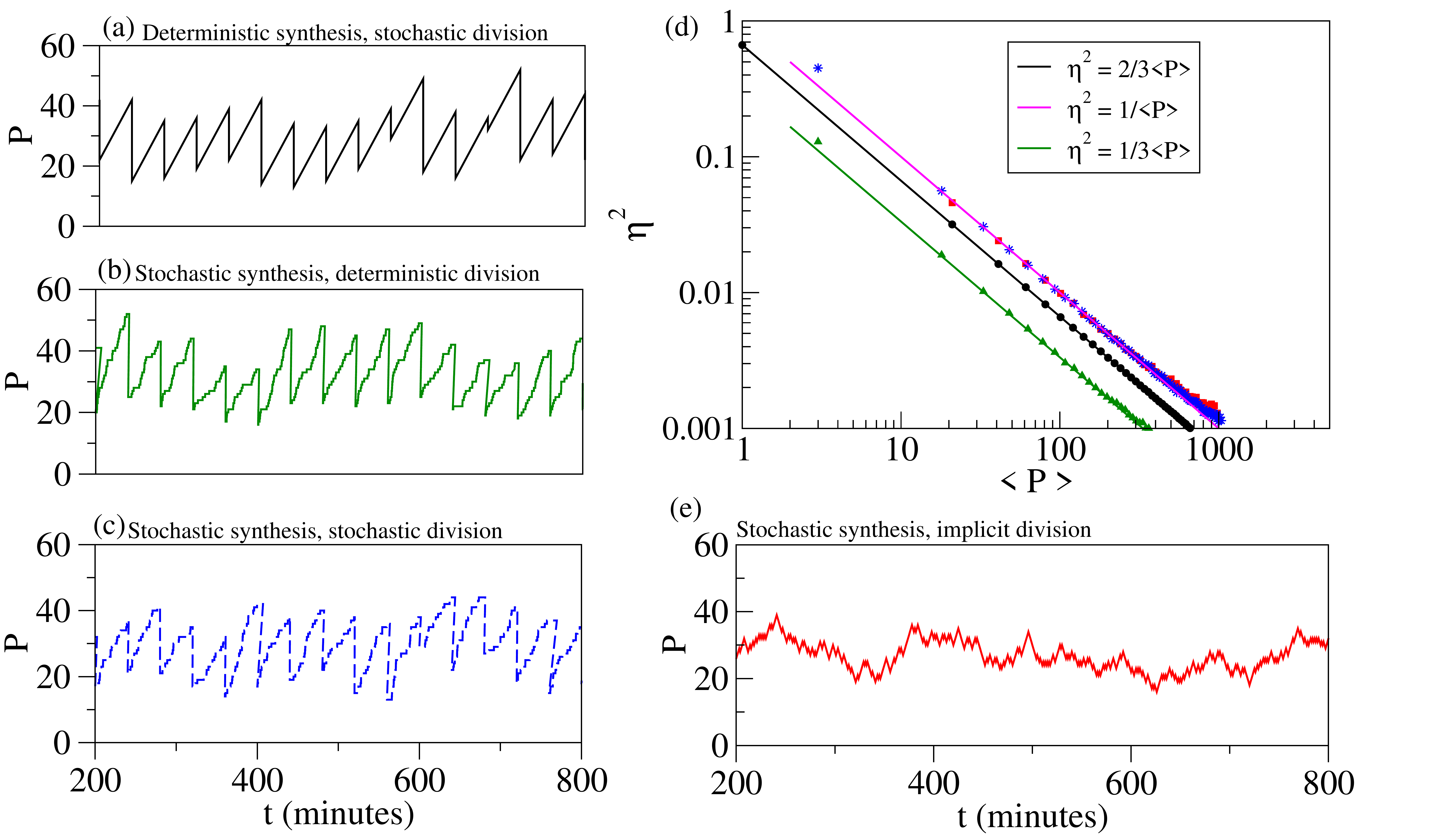}
\caption{Stochastic models of protein synthesis: (a)-(c) Trajectories of the protein copy number from stochastic simulations with stochastic synthesis, cell division, or both, all with cell division modeled explicitly. (d) Noise strength $\eta^2$ as a function of the average protein copy number $\la P\ra$ (varied by varying the synthesis rate $\alpha$) for the different models (for the models with explicit cell division, averages over cell immediately after division are plotted, i.e. $\eta_0^2$ and $\la P_0\ra$). (e) Trajectory of the protein copy number for a model with implicit cell division, i.e.\ where cell division is described by an effective degradation rate $\beta_{\rm eff}$. {\color{black} The corresponding curve in (d) lies on top of the curve for stochastic synthesis and stochastic division. The parameter values used for these plots are $\alpha=0.5/$min, $T= 40$ min, and, in (e), $\beta=\ln 2/T$.  
} 
}
\label{fig:stoch}
\end{figure}

Now let us combine the two sources of stochasticity discussed so far and consider the case where both protein 
synthesis and partitioning are stochastic (Figure \ref{fig:stoch}c). Using again the method of ref.\   
\cite{BrennerPRL07}, we obtain 
\begin{equation}\label{combined}
\la P_0\ra=\alpha T,\qquad \delta P_0^2=\alpha T,\qquad {\rm and}\qquad \eta_0^2=1/\la P_0\ra.
\end{equation}
Two points are noteworthy here: (i) The noise strengths ($\eta^2$) of independent noise sources are additive. In our case, the noise in Eq. \ref{combined} is the sum of the noise components that arise from stochastic partitioning ($2/\la 3 P_0\ra$) and from stochastic synthesis ($1/\la 3P_0\ra$). (ii) The contributions from both sources of noise are of the same order, there is no dominant source of noise in this simple case.

For comparison, we also consider the corresponding model with implicit cell division, i.e.\ a stochastic version of Eq. (\ref{eq:LT}), where the effect of protein dilution through cell growth and division is described by an effective degradation rate $\beta_{\rm eff}=\beta+\ln 2/T$. In this case, we end up with a simple birth-death process, where the number $P$ of copies of our protein of interest increases with constant rate $\alpha$ and decreases with rate $\beta_{\rm eff}P$, described by the following master equation
\bea\label{protsyn}
\frac{\partial \mathcal{P}(P,t)}{\partial t} & = & \alpha\left[ \mathcal{P}(P-1,t) -\mathcal{P}(P,t)\right]\nn\\
 & & +\beta_{\rm eff}\left[ (P+1) \mathcal{P}(P+1,t) 
- P\, \mathcal{P}(P,t) \right],
\label{ptevo}
\eea
where $\mathcal{P}(p,t)$ is the probability to have $P$ proteins at time $t$. The moments of that distribution in the steady state $\la P^n \ra$ can easily be calculated by multiplying the master equation with powers of $P$ and summing over $P$. For this type of model, the protein copy number does not exhibit the periodic behavior seen in the models with explicit cell division, but rather fluctuates around a constant mean value $\la P\ra=\alpha/\beta_{\rm eff}$ in the steady state (Figure \ref{fig:stoch}e). 
These fluctuations are characterized by 
 $\eta^2=1/ \la P\ra$, so the Fano factor is the same as $F_0$ for the case with explicit cell division discussed before. This indicates that using models with implicit cell division (which by the choice of $\beta_{\rm eff}$ are constructed to correctly describe the dynamics of the mean protein number on time scales that are long compared to the generation time $T$) also provide a good description of the fluctuations in such a system.

\section{Bursts of protein synthesis}
\label{sec:bursts}

\begin{figure}
\vspace{0.5cm}
\includegraphics[width=\textwidth]{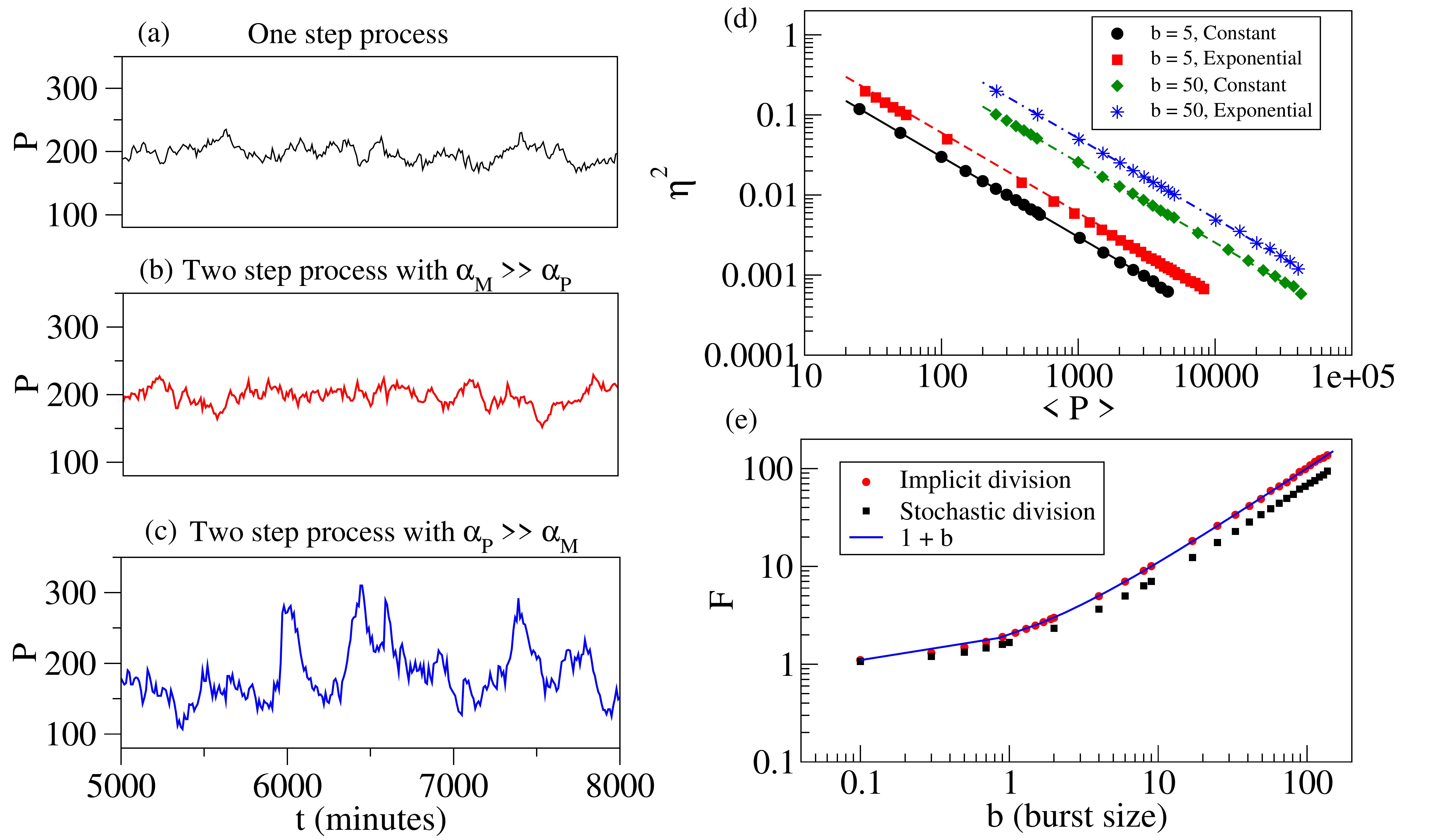}
\caption{Burstiness of protein synthesis: (a)-(c) Trajectories of the protein copy number from stochastic simulations with (a) a one-step model of protein synthesis, (b) a two-step model (transcription and translation) with low burstiness, and (c) a bursty two-step model. All three cases are for implicit cell division and exhibit the same average protein copy number. (d) Noise strength $\eta^2$ for bursty protein synthesis with exponential burst size distribution (as in the two-step models) or with constant burst sizes as a function of the average protein copy number (varied by varying $\alpha_m$). (e) Fano factor for models with  either implicit or explicit cell division as function of the burst size $b$. {\color{black} The parameter values are (a) $\alpha = 2$/min, $\beta_{\rm eff} = 0.01$/min,
(b) $\alpha_p = 0.4$/min, $\beta_p = 0.01$/min, $\alpha_m = 10$/min, $\beta_m = 2/$min, (c) $\alpha_p = 10/$min, $\beta_p = 0.01/$min, $\alpha_m = 0.4$/min, $\beta_m = 2$/min, 
(d) $\beta_p=0.01$/min, and (e) $\beta_p=0.01$/min, $\alpha_m=2$/min, $\beta_m=5$/min, $T=60$ min.}}
\label{fig:burst}
\end{figure}

As discussed in section \ref{sec:mRNA}, the two-step nature of protein synthesis can often be neglected as mRNA levels evolve on faster time scales than protein levels, and therefore the dynamics of mRNA can  be  approximated by its steady state. However, while absorbing the mRNA degrees of freedom into effective protein synthesis results in a correct description of the average protein level, it generally underestimates fluctuations, as it smoothens out the 'bursty' nature of protein synthesis resulting from the two-step process. This was realized first by Berg in 1978 \cite{BergJTB78} and has been studied extensively in recent years, as experimental techniques to count proteins in individual cells were developed 
\cite{CaiNature06,YuSci06,Ozbudak02}. 

To keep the discussion simple, we start with the stochastic version of Eq. (\ref{eq:P_M}), i.e. with a model that describes cell division by an effective protein degradation \cite{OudenaardenPNAS01}.  
The  mRNA part of Eq.(\ref{eq:P_M}),  follows the same dynamics as  the protein in Eq. (\ref{protsyn}) and is thus characterized by the same noise $\eta_M^2 =1/\la M \ra$ with $\la M \ra =\alpha_m/\beta_m$. However the protein number, $P$, behaves differently and is characterized by 
$\la P \ra =\alpha_m\alpha_p/\beta_m\beta_p$ and $\eta_P^2 =(1+b)/\la P \ra$ \cite{OudenaardenPNAS01}, where 
$b= \alpha_p/(\beta_p+\beta_m)\approx \alpha_p/\beta_m$ is called the 'burst size' and describes the average number of proteins synthesized per copy of the mRNA or the amplification of transcription by translation. Experimentally determined burst sizes range between 1 and 10 \cite{YuSci06,xie}. 
The increase in noise can be interpreted as an additional (independent) source of noise that arises from the stochastic amplification of the transcription output by translation. This additional noise is characterized by a noise  strength $b/\la P \ra$ that is added to the noise already present from stochastic protein synthesis and degradation/dilution in the absence of stochastic amplification.

The bursty nature of these processes is shown by cases with low transcription rate: In this case, protein synthesis events are rare (as transcripts are produced infrequently), but multiple copies of the protein are generated in every synthesis event. The increase in fluctuations for the case of bursty synthesis is illustrated in Fig. \ref{fig:burst}, where we plot trajectories for three cases with the same average protein number. In Fig. \ref{fig:burst}(a), protein synthesis is described by a single step with rate $\alpha=\alpha_m\alpha_p/\beta_m\approx \alpha_m b$, in Fig. \ref{fig:burst}(b) and (c) protein synthesis is described as a two-step process. However, while in Fig. \ref{fig:burst}(b) the transcription rate is large and the translation rate is small ($b\simeq 0.2$), the translation rate is large and the transcription rate is small in Fig. \ref{fig:burst}(c), resulting in bursty protein synthesis (with $b\simeq 5$). 

It is worth mentioning here that the bursts on the one hand amplify the noise from transcription, but on the other hand also create additional noise as the size of each burst is a stochastic quantity. To disentangle these two effects, we determine the noise strength $\eta^2$ for a one-step model of protein synthesis, where however $b$ copies of the protein are produced in every synthesis event. In that case, the burst size does not fluctuate, but bursts can still amplify the noise from the one-step synthesis process that mimics transcription. This case can be solved using a modified version of the master equation (\ref{ptevo})\footnote{The first term is replaced by $\alpha \mathcal{P}(P-b,t)\times \Theta(P-b)$, where $\Theta$ is the Heaviside function with $\Theta(P-b)=1$ for $P\geq b$ and $\Theta(P-b)=0$ for $P<b$.} and leads to a noise strength $\eta^2=(1+b)/(2\la P\ra)$ with $\la P\ra=b\times\alpha/\beta_{\rm eff}$. {\color{black} This is exactly half of what we have obtained for exponentially distributed burst sizes in the two-step model} (see also Fig. \ref{fig:burst}(d), where we plot the noise strength for both constant and exponentially distributed burst sizes).\footnote{{\color{black} For constant burst sizes, the values of $b$ must be integers and that the result for a single-step protein synthesis is recovered for  $b=1$, where every transcription event leads to the synthesis of exactly one protein molecule. With stochastic burst sizes, however, $b$ can have non-integer values and the single-step process is recovered by taking the limit $b\to 0$, while keeping $b\times \alpha_m$ constant. } } 
This result indicates that the two effects of bursting contribute equally to the increased noise.


The model discussed so far describes cell division implicitly as an effective protein degradation, but models with explicit stochastic cell division exhibit the same burstiness behavior. This is shown in Fig. \ref{fig:burst}(e), where we plot the Fano factor $F_{P,0}=\eta_{P,0}^2\la P_0\ra$ for a model where both protein and mRNA are divided stochastically between daughter cells as in section \ref{sec:combined}. 
$F_{P,0}$ shows the same dependence on the burst size except for a different prefactor of the linear term ($\approx \ln 2$), which arises from the fact that averages are taken over slightly different populations (over cells immediately after division vs.\ over age-less cells representing an average over the division cycle). 

Finally, we want to mention that burstiness can also arise from other physical processes than from multiple translations of a transcript. For example, bursts have been demonstrated experimentally to occur on the level of transcription \cite{Golding}, which can be interpreted as resulting from the stochastic switching of the gene between two activity states (transcription 'on' or 'off').  The molecular origin of these activity states remains however unclear,\footnote{In eukaryotic systems, they are believed to mostly reflect different states of the chromatin structure.} although several mechanisms have been proposed (e.g. states of chromosome structures, binding/unbinding of transcription factors, etc.  \cite{Mitarai,Walczak05}). 
In a genome-wide study, the Fano factors for mRNA were found to range mostly between 1 and 2, larger than what is expected for a single-step (Poisson) synthesis, but not much larger \cite{xie}.


\section{Extrinsic noise}
\label{sec:extrinsic}

So far, we have discussed intrinsic noise in gene expression, i.e. noise that is specific to a particular gene or protein and results from the inherent stochasticity of the synthesis and degradation of that protein. As we have seen,  a characteristic property of intrinsic noise is its scaling proportional to the inverse of the average protein number in the cell. We now turn to extrinsic noise, fluctuations of cellular parameters that affect all genes/proteins in a cell. Such noise has first been demonstrated by a study of the correlations between the reporter proteins expressed from two copies of the same operon \cite{ElowitzSci02}. For highly abundant proteins, intrinsic noise becomes negligible and the extrinsic component of the noise, which does not depend on protein abundance, is dominant with fluctuations of about 30\% in the protein concentration as shown by a study of a library of fluorescent reporter proteins \cite{xie}. There are many possible sources of extrinsic noise such as fluctuations in the concentrations of essential components of the transcription and translation machinery or mRNA degradation enzymes (RNA polymerases, ribosomes, RNases). Here we consider two effects that should be present even if such fluctuations are suppressed by feedback mechanisms for the synthesis of these machines: cell-to-cell variations arising from different cell ages in a population (section \ref{sec:6}) and effects due to fluctuations in the growth rate (section \ref{sec:growth}). 

{\color{black}
We note that another definition of extrinsic and intrinsic noise has been given in ref.\  
\cite{paulsson04}. There, the distinction between extrinsic and intrinsic noise is not based on distinguishing a specific genetic system and its environment, which affects different genes in the same way, but on the dependence of the noise on the average protein number. One component of the noise  exhibits the characteristic $1/\la P \ra$ behavior and is classified as intrinsic, while the component of the noise that does not exhibit this behavior and depends on the fluctuation of a variable that influences the protein synthesis rate is be classified as extrinsic. The two cases we consider here are extrinsic according to both definitions, but based on the definition of ref. \cite{paulsson04}, one could, for example, consider the noise from transcription as extrinsic to translation. } 

\subsection{Effects of the division cycle}
\label{sec:6}


In section \ref{sec:1}, we have seen that the protein concentration varies systematically over the course of a division cycle. In a population of non-synchronized cells, this age-dependence of the protein content is observed as a  cell-to-cell variation that forms part of the extrinsic noise. 
To study the effect of age-dependent protein content and to estimate what part of the extrinsic noise can be understood from such deterministic variation, we now determine the distributions of the protein number and concentration over the division cycle. As for the average protein number calculated in section \ref{sec:1}, we have to take the age distribution of the experimental culture into account. We consider again the case of a single lineage and of an exponentially growing population, i.e. a constant and an exponential age distribution as given in Eq.~\ref{age_distr}.
We denote the resulting protein copy number distributions by $\Phi(\concp)$ and $\Psi(\concp)$. They can be calculated by inverting the time dependence of $P(t)$ and using the inverse relation $t(P)$ for a transformation of variable in the age distribution, see appendix \ref{app:distribution}. 

\begin{figure}[tb]
 \begin{centering}
\includegraphics[width=1\textwidth]{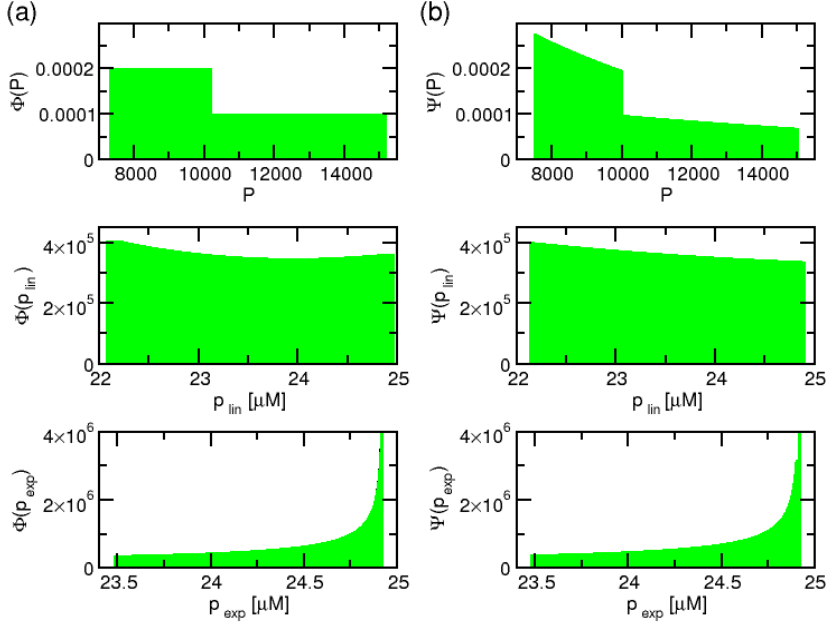}
 \caption{Distributions for protein number and concentrations as arising from the deterministic variation over the division cycle. (a) Distributions for the protein number $\Phi(\concp)$ and concentration $\Phi(\xlin)$ and $\Phi(\xexp)$ (for the case of linear and exponential volume growth, respectively) for a single lineage. 
(b) Distributions for the protein number $\Psi(\concp)$ and concentrations $\Psi(\xxlin)$ and $\Psi(\xxexp)$ for an exponentially growing cell population with age distribution $\agedistr$.  The parameters are as in Fig.~\ref{fig1}. \label{fig3}}
 \end{centering}
 \end{figure}

The distributions for both types of cell culture are presented in Fig.~\ref{fig3}. Panel (a) shows the distributions of protein number and concentration for a single lineage, $\Phi(\concp)$ and $\Phi(\concx)$, respectively. The concentration distribution was determined for both linear and exponential volume growth. the distribution of the protein copy number, $\Phi(\concp)$ (top panel), exhibits two flat plateaus. The probability to find a protein number $P<\concp(t=\tx)$ that is seen prior to the replication time $\tx$ is twice as high as for a protein number that corresponds to larger times, $\concp(t>\tx)$, as the synthesis rate doubles at time $\tx$. 

For the concentration subject to linear volume growth (middle panel), $\Phi(\xlin)$ is almost flat with a minimum for intermediate concentrations.
In the case of exponential volume growth (bottom panel),  $\Phi(\xexp)$, which is quite flat for small concentrations, rises sharply towards the maximum concentration.

It is worth noting that while the protein copy number exhibits a broad distribution over a two-fold range, defined by the copy numbers directly before and after cell division, the range over which the concentration varies is much smaller: The maximal concentration is only $\approx$13\% larger than the minimal concentration for linear volume growth and even less ($\approx$ 6\%) for exponential volume growth). 

Figure \ref{fig3} (b) shows the corresponding results for an exponentially growing culture (with an exponential age distribution). 
The distribution for the protein number (top panel), $\Psi(\concp)$, still exhibits two plateaus, which are tilted towards smaller values of $\concp$ as the age distribution gives more weight to younger cells. 
The distributions of the concentrations (middle and lower panel), $\Psi(\xxlin)$ and $\Psi(\xxexp)$, are not radically altered by the change in age distribution.

\begin{figure}[tb]
\begin{centering}
\includegraphics[width=.7\textwidth]{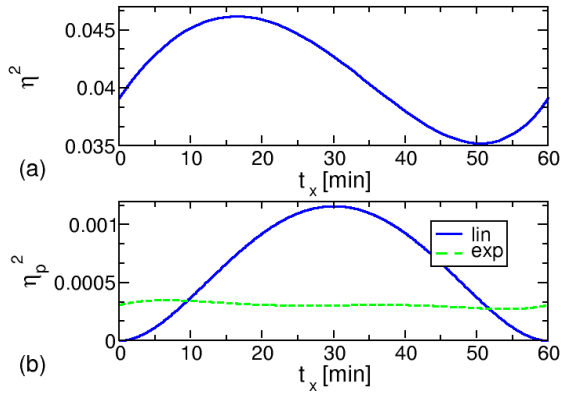}
\caption{Noise parameter $\eta^2$ arising from deterministic variations over the division cycle for (a) the protein number $\concp$ and (b) for the protein concentration, $\etaxlin$ and $\etaxexp$ (with linear and exponential volume growth) as functions of the replication time $\tx$.  \label{fig2}}
\end{centering}
\end{figure}

Next, we determine the noise parameter that characterizes the variation over the division cycle, which in analogy to Eq.\ {\ref{def_eta}} can be defined as 
\begin{eqnarray}
\eta^2=\delta\concp /\langle \concp\rangle^2=\frac{\int_0^T (\concp(t)-\langle \concp\rangle)^2 \agedistr dt}{\langle \concp \rangle^2}.
\end{eqnarray}
for the protein copy number and likewise, $
\etaxlin=\delta\xxlin/\langle \xxlin\rangle^2$ or $\etaxexp=\delta\xxexp/\langle \xxexp\rangle^2$, for the concentration (for linear and exponential volume growth, respectively). 
One parameter that affects the extent of this deterministic cell-to-cell variation is the replication time $\tx$, which depends on the genomic location of the gene of interest relative to the origin of replication. In Fig.~\ref{fig2}, we show the noise parameters for the protein copy number and the concentration as functions of $\tx$ in the range of $0\leq\tx \leq T$.  

We first note that these noise parameters only depend on $T$ and $\tx$, or, more precisely, on their ratio  $\tx/T$. Specifically they are independent of the protein synthesis rate $\alpha$ and, in case of the concentrations, the initial cell volume $V_0$. Therefore, this contribution to the observed noise does {\color{black} not} decrease with increasing protein concentration and does not become negligible for abundant proteins. 
However, the overall contribution of the division cycle to the noise is relatively small. For the protein concentration, the noise $\etaxlin$ in the case of linear growth is on the order of $0.001$ (solid line in Fig. \ref{fig2}b), and for exponential growth $\etaxexp\leq 5\cdot 10^{-4}$ for all values of $\tx$. These values correspond to 2-3 \% variation of the concentration and are considerably smaller than the observed extrinsic noise, which is of the order of $\eta^2\simeq 0.1$ \cite{xie}. 
For the protein copy number, which varies over a wider range, the effect of the division cycle is more pronounced and varies by about one fourth of its mean over different values of $\tx$. However, even here, the absolute value of $\eta^2$, being on the order of 0.04, remains rather small. We can thus conclude that, while the division cycle contributes to the observed extrinsic noise, other sources of extrinsic noise are more dominant.

\subsection{Fluctuations of the growth rate}
\label{sec:growth}

The  fluorescent reporter protein library study of Taniguchi et al.\  mentioned above \cite{xie} showed that abundant proteins exhibit extrinsic noise that does not display the inverse scaling with the mean protein concentration. The same study also revealed some additional characteristics of that noise:  In particular, (i) there are correlations between the noise of different extrinsic proteins, a defining feature of extrinsic noise \cite{ElowitzSci02}, and (ii) the extrinsic fluctuations are slow, with variations in the protein concentration over timescales longer than the generation time \cite{xie}. Moreover, they come together with substantial fluctuations of the generation time. We therefore  ask now whether fluctuations in the growth rate may substantially contribute to the observed extrinsic noise. 

For an estimate of the effect of a fluctuating growth rate, we make the assumption that while the doubling time fluctuates slowly, the protein synthesis rate per cell volume $\alpha/V$ remains approximately constant. This condition is (approximately) satisfied by the population average of the synthesis rate as a function of growth rate when the growth rate is systematically varied by using different growth media \cite{Klumpp09}. It basically means that changes of the growth conditions, while affecting the synthesis rate of protein numbers, do not affect the rate of synthesis of protein concentration. Only the effective degradation is changed when the growth rate changes. Under balanced growth, this constancy is the result of the combination of several factors (such as the availability of RNA polymerases and ribosomes, the gene copy number etc. \cite{Klumpp09,Klumpp08}) that do change, but in such a way that their combined effect cancels out (with the exception of conditions of very slow growth) \cite{Klumpp09}. Obviously, it is not clear that this assumption holds for slowly varying growth rates in individual cells; in principle, all factors that contribute to the growth-rate dependence of protein concentrations could vary in a mutually  independent fashion, but we can consider the case where they vary together as one that provides a lower limit for the resulting noise. 
With a deterministic description of protein synthesis, we obtain $p=(\alpha/V)\times T/\ln 2$, so fluctuations of $T$ are directly carried over into fluctuations of the protein concentration $p$. If $T$ fluctuates by some time $\Delta T$ (of about 10--25 \% of the doubling time), $p$ will fluctuate by $\Delta p=\alpha \Delta T/(V \ln 2)$ or also about 10--25 \%, as $\Delta p/p=\Delta T/T$. This would correspond to a noise parameter $\eta^2$ of 0.01--0.08.
While this simple estimate is certainly not an accurate description of such global noise, it clearly indicates that fluctuations in the growth rate can lead to noise in protein concentrations of the order of the observed extrinsic noise \cite{xie}.

\section{Concluding remarks}
\label{sec:concl}

In this article, we have discussed several ways of describing gene expression with deterministic or stochastic models. Deterministic models that explicitly describe cell division, gene duplication, and  volume growth provide a detailed description of the dynamics over both short and long time scales (compared to the doubling time). We have shown that the results depend generally on specific details of the model such as how volume growth is implemented and the age structure of the population over which averages are taken. Fortunately, however, these differences are not dramatic. Moreover, a mean-field-like approximation that describes protein synthesis by an effective rate per volume given by the average gene copy number and the average cell volume provides a good approximation that averages over the detailed dynamics within the division cycle. Nevertheless, it is worth keeping in mind that there are all these subtle effects as well as to carefully distinguish different normalizations of protein amounts or synthesis rates such as per gene (e.g.\ $\alpha$), per cell ($\alpha \la g\ra$) and per volume ($\alpha \la g\ra/\la V\ra$). This is particularly important for studies that address the coupling of gene expression and global cellular physiology, where quantities such as the average gene copy number and the average volume per cell may change \cite{Klumpp09}.  

With respect to the fluctuations around this average behavior, we have compared several simple models to disentangle the contributions of different sources of noise. This comparison shows that the noise contributions from sources such as stochastic protein synthesis or degradation and stochastic partitioning during cell division are all of the same order and that there is no single dominant noise source, except when protein synthesis is pronouncedly bursty. The burstiness of protein synthesis is the largest contribution to the noise (with a Fano factor $\approx b$, while the other noise sources have Fano factors of fractions of 1). If $b$ is large, this is clearly dominant, and one could neglect all other sources of noise.  The study of Taniguchi et al. \cite{xie}, however, indicates that typical values of $b$ for many low-abundance proteins are in the range 1--10 and thus are not necessarily very dominant. {\color{black} In many cases, a realistic description of the dynamics of expression of low-abundance proteins will therefore need to include all these sources of noise. }

For intermediate-abundance to high-abundance proteins (with $\la P\ra >20$), the noise is dominated by extrinsic noise \cite{xie}. Here we have considered two sources of extrinsic noise: We have shown that the deterministic contribution from systematic variation over the division cycle is rather small (even for the protein copy number, but in particular for the concentration), while fluctuations in the growth rate can be expected to give a larger contribution. These results suggest that a model that incorporates the burstiness of protein synthesis and fluctuations in the growth rate might provide a minimal description of stochastic effects in gene expression that is able to describe both intrinsic and extrinsic components of the noise.

\begin{acknowledgements}
The authors would like to thank Angelo Valleriani for stimulating discussions during the course of this work.
\end{acknowledgements}

\appendix
\section{Appendix}

\subsection{Typical values of the parameters}
\label{app:param}

{\color{black} Estimates of  typical parameter values in the model organism {\it E.\ coli} are summarized in Table \ref{tab:param}. Most of these can, for example, be estimated from the data of ref.\ \cite{xie}. A few of them require additional comments: (i) In {\it E.\ coli} proteins are typically stable, i.e.\ $\beta_p\approx 0$. So far, no complete survey of protein stability has been made, but the total cellular protein mass was found to 
be stable \cite{NathJBC70} and early proteomics studies (2d-gels) also indicated that almost all proteins covered by their approach were stable \cite{ReehMGG79}. Nevertheless, some proteins are known to be unstable and, in these cases, $\beta_p$ can be of the order of 1 min$^{-1}$. (ii) Genes are typically present as a single copy in the genome. This means that the gene  copy number per cell is 1 before the gene is replicated and 2 after replication. Average gene copy numbers are between 1 and 2, except at fast growth with doubling times $T< 60$ min, where rounds of DNA replication overlap and the gene copy numbers can be larger \cite{CHelmstetter,Bremer1996}. (iii)  The cell volume doubles over the division cycle and its average value depends on  the growth conditions \cite{Bremer1996}. The value given in the table should be taken as an order or magnitude estimate. 

\begin{table}[b]
\caption{Typical parameter values for {\it E.\ coli} cells}
\begin{center}
\begin{tabular}{l |c| l | l}
parameter & symbol & typical range & comments\\   \hline
transcription rate & $\alpha_m$  & 0.1--10 min$^{-1}$ &   \\ 
mRNA degradation rate & $\beta_m$ & 0.2--2 min$^{-1}$ &    \\ 
translation rate & $\alpha_p$ &  1--10 min$^{-1}$ &  \\
protein degradation rate & $\beta_p$ & $\approx 0$ & see text \\ \hline
gene copy number & $g$ & 1--2& see text \\
division time & $T$ & 20 min -- hours & \\ 
average cell volume & $\langle V_0 \rangle$ & $\approx 1 \mu$m$^3$ & see text\\ \hline
effective synthesis rate &  $\alpha$ & 0.1--500 min$^{-1}$ &   $ =\alpha_m \alpha_p g/\beta_m$ \\
burstiness & $b$ & $1-50$ & $ \approx \alpha_p/ \beta_m$ \\
effective degradation rate & $\beta_{\rm eff}$ & $\sim 0.01$ min$^{-1}$ & $ = \beta_p +\ln 2/T\ $ \\ 
\end{tabular}
\end{center}
\label{tab:param}
\end{table}%

}  


\subsection{Models with stochastic protein synthesis and stochastic division.}\label{app:method}

A general method for solving processes involving different rules of protein synthesis and cell division 
has been described in ref.\ \cite{BrennerPRL07}. This method allows us in most of the cases to find averages 
and standard deviation of the protein number. We will describe the method briefly here following 
\cite{BrennerPRL07}. Let $P_n$ be the protein content in the $n^{th}$ generation immediately after 
the cell division. Let $\lambda_n$ be the amount of protein produced and accumulated till the cell 
division time in generation $n$ and $q_n$ be the fraction of protein inherited by the daughter cell 
at the time cell division. Then one can write
\bea
P_{n+1} = q_n~ (P_n + \lambda_n).
\eea
The protein generation as well as division can be taken from some distributions. If these distributions admit
finite moments then in the steady-state the distributions of $\lambda$ and $q$ become independent and hence one can write
\bea
\langle P^k \rangle = \langle q^k \rangle~\langle (P+\lambda)^k \rangle. 
\label{protmome}
\eea
From here one can get all the moments for $P$,  in particular $\langle P \rangle = \langle \lambda \rangle$.
Let us consider an example where we add protein with rate $\alpha$ in between every two cell divisions and where the
protein number is divided deterministically into half at every cell division after every $T$ time. In this case 
the synthesis of protein follows a binomial distribution giving $\langle \lambda \rangle =\delta\lambda^2 = \alpha T$  and the
division fraction {\color{black} is given by a delta function $\delta(q-1/2)$} with $ \langle q \rangle = 1/2$ and  $\langle q^2 \rangle - \langle q \rangle^2 = 0$.
Thus Eq.(\ref{protmome}) gives $\langle P \rangle = \langle \lambda \rangle$ and $\langle P^2 \rangle = \frac{1}{3}\ 
(2\langle \lambda \rangle^2 + \langle \lambda^2 \rangle)$. After some algebra one finds 
$\eta^2 = \frac{\langle P^2 \rangle- \langle P \rangle^2}{\langle P^2 \rangle}\ = \frac{1}{3\langle P_0 \rangle}\ $
which is one of the cases discussed in the main text.


\subsection{Distribution of protein number and concentration due to variation over the division cycle\label{app:distribution}}

The distribution of the protein number discussed in section \ref{sec:6} is obtained by inverting the time-dependence of the 
protein copy number, $P(t)$ to obtain $t(P)$ and a transformation of variables in the age distribution from $t$ to $P$, which leads to 
\begin{eqnarray}\label{psiage}
\Psi(\concp)=\left(\frac{d}{d\concp}t(\concp)\right)\phi(t(\concp)).
\end{eqnarray} 
Specifically, for the constant age distribution that describes averages over a single lineage, this leads to
$\Phi(\concp)=\frac{d}{d\concp} t(\concp)$. 
As a consequence, the result for an arbitrary age distribution can be rewritten as 
\begin{eqnarray}\label{psiage1}
\Psi(\concp)=\Phi(\concp)\phi(t(\concp)),
\end{eqnarray} 
i.~e., the distribution of protein number in a single lineage weighted with the age distribution of the corresponding inverse. 

The distributions for the concentrations are obtained in an analogous fashion, but the calculation is technically more involved as the concentration is not a monotonic function of time (see, e.g. Fig.~\ref{fig1}). We thus split the functions $\xlin$ and $\xexp$ into piecewise monotonic functions and determine the distributions for these separately. The concentration for linear cell growth, $\xlin$, is monotonic in the intervals $[0,\tx]$ and $[\tx,T]$, and for $\xexp$ we have three intervals $[0,\tx]$, $[\tx, \tmax]$ and $[\tmax,T]$, where $\tmax$ is the time where $\xexp$ is maximal.
The complete distributions $\Phi(\xlin)$ and $\Phi(\xexp)$ are then obtained by adding up the distributions from the respective intervals. 
The distributions for the concentrations $\Psi(\concx)$, are again obtained for the corresponding intervals, weighted with the age distribution and summed up to yield the full distribution.


\end{document}